\documentclass[%
 aps,prb,rsi,amsmath,amssymb,superscriptaddress,reprint,
]{revtex4-2}

\usepackage{graphicx}
\usepackage{dcolumn}
\usepackage{bm}
\usepackage{lipsum}

\begin{document}

\title{Easy-plane spin Hall nano-oscillators as spiking neurons for neuromorphic computing}

\author{Danijela Markovi\'c}\thanks{Author to whom correspondance should be addressed. Electronic mail : danijela.markovic@cnrs-thales.fr}
 \affiliation{Unit\'e Mixte de Physique CNRS/ Thales, Universit\'e Paris-Saclay, 91767 Palaiseau, France}

  \author{Matthew W. Daniels}
 \affiliation{Physical Measurement Laboratory, National Institute of Standards and Technology, Gaithersburg, Maryland, USA}
 
\author{Pankaj Sethi}
\affiliation{Unit\'e Mixte de Physique CNRS/ Thales, Universit\'e Paris-Saclay, 91767 Palaiseau, France}
 
\author{Andrew D. Kent}
 \affiliation{Center for Quantum Phenomena, Department of Physics, New York University, New York, USA}
 
\author{Mark D. Stiles}
 \affiliation{Physical Measurement Laboratory, National Institute of Standards and Technology, Gaithersburg, Maryland, USA}
 
\author{Julie Grollier}
\affiliation{Unit\'e Mixte de Physique CNRS/ Thales, Universit\'e Paris-Saclay, 91767 Palaiseau, France}

\begin{abstract}
We show analytically using a macrospin approximation that easy-plane spin Hall nano-oscillators excited by a spin-current polarized perpendicularly to the easy-plane have phase dynamics analogous to that of Josephson junctions. Similarly to Josephson junctions, they can reproduce the spiking behavior of biological neurons that is appropriate for neuromorphic computing. We perform micromagnetic simulations of such oscillators realized in the nano-constriction geometry and show that the easy-plane spiking dynamics is preserved in an experimentally feasible architecture.
Finally we simulate two elementary neural network blocks that implement operations essential for neuromorphic computing. First, we show that output spikes energies from two neurons can be summed and injected into a following layer neuron and second, we demonstrate that outputs can be multiplied by synaptic weights implemented by locally modifying the anisotropy.

\end{abstract}

\maketitle

Spintronic nano-oscillators can emulate neurons: their nonlinear dynamics has already enabled multiple demonstrations of supervised learning~\cite{Mizrahi2018, Torrejon2017, Romera2017a, Markovic2019}. However, spintronic devices have not yet produced the spiking behavior of biological neurons.

Spiking dynamics is interesting for neuromorphic computing for several reasons: it allows for particularly energy-efficient encoding of information, but it could also allow for the implementation of local learning rules such as spike timing-dependent plasticity (STDP) that enables some forms of unsupervised learning.

On the other hand, several groups are pursuing Josephson junctions as spiking neurons in neuromorphic computing schemes. Indeed, the superconducting phase in Josephson junctions can be made to oscillate so predictably that they consitute the realization of the volt within the Syst\`{e}me international d'unit\'{e}s (SI). They can also operate in a spiking regime in which they make single cycles in their phase~\cite{Crotty2010, Segall2017}.  The chaotic physics of these systems can give rise to rich and useful neural dynamics, but there the picosecond timescale of the phase spikes~\cite{Russek2016} and the low-temperature setup required for superconducting physics both make integration of Josephson junction neurons with other computing technologies challenging.

Khymyn et al.~have recently showed that spiking dynamics similar to that of Josephson junctions could be obtained in spin Hall nano-oscillators (SHNOs) based on antiferromagnetic materials ~\cite{Khymyn2017}. These devices could produce voltage spikes at terahertz rates provided the injected spin-current is polarized perpendicularly to the easy-plane. Their magnetization undergoes a precession in the easy-plane with a phase that can be described by equations analogous to those of the superconducting phase in Josephson junctions~\cite{cheng2015ultrafast}. Similarly to the voltage spikes that Josephson junctions emit above the critical supercurrent~\cite{Crotty2010, Segall2017, Russek2016}, easy-plane antiferromagnetic oscillators can emit voltage spikes above the critical spin current density. This result is promising for its applications in terahertz generation, neuromorphic computing,~\cite{Khymyn2018a}, and macroscopic antiferromagnetic qubits~\cite{Takei2017}. However, it is challenging to realize these devices because of the difficulty of growing easy-plane antiferromagnetic materials and controlling their domain structure. Attempts to address this difficulty have led to the discovery that similar dynamics can be obtained in synthetic antiferromagnetic junctions~\cite{Liu2020}.

In this work we demonstrate spiking behavior in an analogous architecture with easy-plane ferromagnetic materials, which are more readily mastered experimentally. Besides their capacity to spike, easy-plane ferromagnetic oscillators are fundamentally interesting for their characteristic oscillation properties emerging from the circular precession as opposed to the elliptical precession more common in easy-axis oscillators. Circular precession at angular frequency $\omega$ can eliminate the emission at $2\omega$ that increases the effective damping~\cite{Divinskiy2019}. As a consequence, easy-plane oscillators should have both lower frequency and damping than the easy-axis oscillators.

\begin{figure}
\includegraphics[scale=0.8]{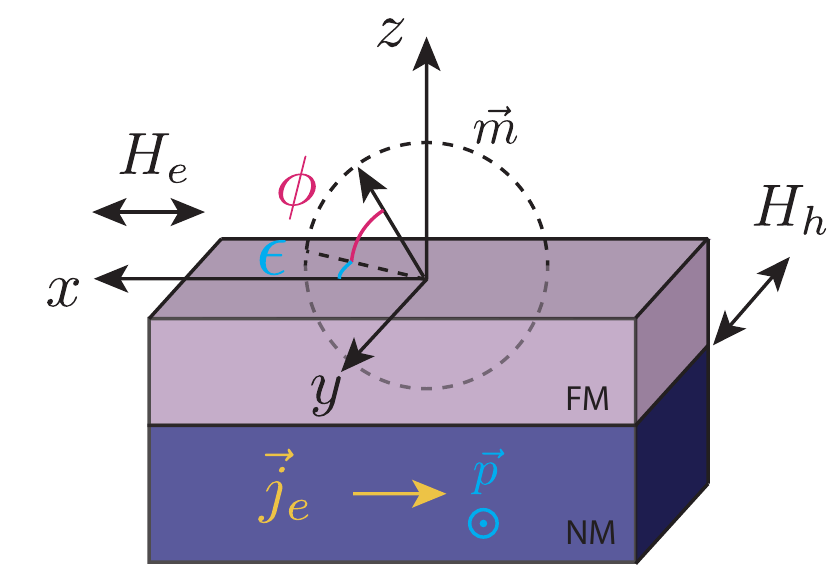}
\caption{Schematic of an easy-plane spin Hall nano-oscillator. The bottom layer is a normal heavy metal (NM) that polarizes the electrical current $j_e$ in spin. The top layer is a ferromagnet (FM) with a net easy-plane (the $xz$-plane) perpendicular to the polarization $\vec{p}$ of the injected spins. The magnetization $\vec{m}$ precesses in the $xz$-plane with phase $\phi$ and small out-of-plane tilt $\epsilon$.}
\label{anisotropie}
\end{figure}

In the first part of this article we show using the macrospin approximation that Josephson dynamics can be obtained in an easy-plane ferromagnetic spin Hall architecture. We explore the spiking dynamics of this model and show that a spiking macrospin neuron can trigger the activation of a second macrospin neuron via the spin wave emissions resulting from a spike event. In the second part we perform micromagnetic simulations and show that such dynamics is preserved for experimentally feasible nano-constriction oscillator architecture~\cite{Demidov2015}. In the third part we show that these oscillators can be coupled in two-dimensional arrays through synapses whose weights can be controlled by locally modifying the anisotropy. We also show that the nano-constriction geometry allows for a multi-input neuron. We close by discussing prospects for neuromorphic computing.

\section{Macrospin model}
\label{sec:macrospin}

To motivate the essential physics of the easy-plane ferromagnetic neuron, we show how an appropriately chosen system can exhibit dynamics described by the same damped driven pendulum physics that governs the Josephson equation. We consider a bilayer, shown in Fig.~\ref{anisotropie}, composed of a low damping ferromagnet with perpendicular magnetic anisotropy (PMA) on top of a non-magnetic heavy metal with large spin-orbit coupling which polarizes the spin of the electrical current through the spin Hall effect. In the macrospin approximation, the dynamics of the scaled ferromagnet magnetization, $\bm{m}=\bm{M}/M_s$, where $M_s$ is the saturation magnetization, follows the Landau-Liftshitz-Gilbert (LLG) equation
\begin{equation}
    \dot{\bm{m}} = - \gamma_0 \bm{m} \times \bm{H}_\text{eff} + \alpha \bm{m} \times \dot{\bm{m}} + \tau \bm{m} \times (\bm{m} \times \bm{p}),
    \label{eq:LLG}
\end{equation}
where $\tau = \sigma j$ is the spin torque, $j$ is the current density, $\alpha$ is the Gilbert damping, $\bm{p} = \bm{e_y}$ is the spin current polarization direction, $\gamma_0=\mu_0\gamma $, $\mu_0$ is the permeability of free space, and $\gamma$ is the gyromagnetic ratio. Since the spins that are injected from the heavy metal are along the $y$-direction, we tune the shape and magnetocrystalline anisotropies to create a net easy $xz$-plane. This can be accomplished by tuning the PMA to roughly cancel the out-of-plane shape anisotropy and elongating the slab in Fig.~\ref{anisotropie} along the $x$-direction to create a net hard axis along $y$ and a weak easy axis along $x$ ~\cite{Legrand2019}.  We model this anisotropy with an effective magnetic field
\begin{equation}
\bm{H}_\text{eff} = H_e \bm{e}_x - H_h \bm{e}_y,
\end{equation}
with characteristic easy- and hard-axis anisotropy frequencies $\omega_e = \gamma_0 H_e$ and $\omega_h = \gamma_0 H_h$. The LLG equation above underpins both the macrospin and micromagnetics simulations throughout the remainder of this paper. In the next section, we reduce this equation to an effective equation of motion in $\phi = \text{arctan}(m_z/m_x)$, and show that it maps onto several other systems of physical interest. Then in Sec.~\ref{sec:macrospin-simulation} we show that simple simulations of a few coupled macrospins can capture the physics of spike transmission from neuron to neuron through a magnetic medium, motivating the full-scale micromagnetic simulations developped in the rest of the paper.

\subsection{Mapping to the Josephson equation}
\label{sec:mapping}
For easy-axis anisotropy much smaller than hard-axis anisotropy, $\omega_e \ll \omega_h$, to linear order in $\omega_e/\omega_h$ we find from Eq.~\eqref{eq:LLG} that the precession phase, 
\begin{equation*}
\phi = \arctan\left(\frac{m_z}{m_x}\right),
\end{equation*} 
has the equation of motion
\begin{equation}
    \frac{1}{\omega_h}\ddot{\phi} + \alpha \dot{\phi}+\frac{\omega_e}{2}\left(1+\frac{2}{\omega_h^2} \dot{\phi}^2\right) \sin 2 \phi = \sigma j .\label{eq:phi-ddot}
\end{equation}
Under the change of variable $\delta = 2\phi$, and neglecting for now the $\dot\phi^2$ term, this is analogous to that of a superconducting phase $\delta$ of a Josephson junction
\begin{equation}
  \ddot{\delta} + \frac{1}{RC} \dot{\delta}+\frac{2 e I_c}{\hbar C}\sin \delta = \frac{2 e}{I_c}I,\label{eq:jj}
\end{equation}
with R, C, and $I_c$ the resistance, capacitance and critical current, or equivalently the equation of motion for a damped driven pendulum. Also of this form are the dynamics of the precessional phase $\phi$ of an antiferromagnetic spin Hall oscillator
\begin{equation}
    \frac{1}{\omega_\mathrm{ex}}\ddot{\phi} + \alpha \dot{\phi}+\frac{\omega_e}{2} \sin 2 \phi = \sigma j.\label{eq:afm}
\end{equation}
The hard axis anisotropy $\omega_h$ in the ferromagnet gives the inertia to the system, similarly to the exchange coupling $\omega_\text{ex}$ in the antiferromagnet and the capacitance $C$ in the Josephson junction. As in the antiferromagnetic system, the second-order inertial dynamics arises by integrating out the small out-of-plane magnetization. 

In the case of the ferromagnetic easy-plane SHNO there exists an additional $\frac{\omega_e}{\omega_h^2} \dot{\phi}^2 \sin 2\phi$ term that cannot be neglected. Nevertheless, in the limit $\omega_e \ll \omega_h$, the precession dynamics is dominated by the hard-axis anisotropy, $\dot{\phi} \simeq \omega_h$, such that
\begin{equation}
    \frac{\omega_e}{\omega_h^2} \dot{\phi}^2 \simeq \omega_e.
\end{equation}
This additional term is of the same order as the other $\sin 2 \phi$ term and presents a velocity-dependent contribution to the restoring force. That the restoring force differs by a factor of two between the rest and precessional states of the oscillator may quantitatively complicate the post-spike relaxation of the system, but the spiking behavior we observe in simulation is qualitatively similar to that of Eqs.~\eqref{eq:jj} and \eqref{eq:afm}.

\subsection{Simulation of spiking behavior}
\label{sec:macrospin-simulation}

In analogy to the superconducting phase across Josephson junctions obeying Eq.~\eqref{eq:jj}, we identify $2\pi$ phase slips of the in-plane angle $\phi$ with neural spiking events, that is, a $2\pi$ rotation in $\phi$ corresponds to a single spike.
In order to generate coordinated spiking behavior in one of these oscillators, we require that the driving torque $\tau$ sit near the spiking threshold. For $\ddot\phi = \dot\phi = 0$, Eq.~\eqref{eq:phi-ddot} exhibits a fixed point at $\phi^* = \text{arcsin}[2\tau/\omega_e]/2$, which has a real solution only when $2\tau \leq \omega_e$, and has a vertical asymptote at $\pi/4$. This value of $\pi/4$ is the critical angle beyond which the driving force will generate a full $2\pi$ rotation of the system. It is the analogous to the $\pi/2$ angle of a damped driven pendulum with respect to the direction of gravity. By tuning the current in our oscillator to move the fixed point as close to $\pi/4$ as possible, we can make the neuron's spiking dynamics arbitrarily sensitive to incoming torques, such as those arising from spin waves due to another neuron's spike emission.

From a macrospin perspective, holding $\phi^*$ close to $\pi/4$ is not alone sufficient to generate coherent spiking. If $\phi^*$ is near $\pi/4$ but $\alpha$ is too small, then a single spike event will take the system into an autooscillation regime and a rest state can never be recovered. This can be understood by imagining the neighborhood around $\phi^*$ as a simple harmonic oscillator. If $\phi^* = \pi/4-\epsilon$ and $\alpha$ is too small, then the system will have an underdamped response when it returns to $\phi^*$ after a spike event. If the amplitude of the underdamped oscillations are greater than or equal to $\epsilon$ then the system will not settle into $\phi^*$ but instead pass the critical threshold $\phi=\pi/4$ and undergo another spike event. Since we want $\epsilon$ arbitrarily small, we use a large damping value of 0.5  in the macrospin simulations to ensure an overdamped return to equilibrium (see Appendix~\ref{app:macrospin-details} for more details). In a real system however, the Gilbert damping should be small for efficiently communicating angular momentum between neurons. A solution is to use large synapses into which energy can be rapidly evacuated over a single spike period as a way to present a large \emph{effective} damping for these oscillator neurons, as shown in Section II. 

\begin{figure}
    \centering
    \includegraphics[width=0.8\columnwidth]{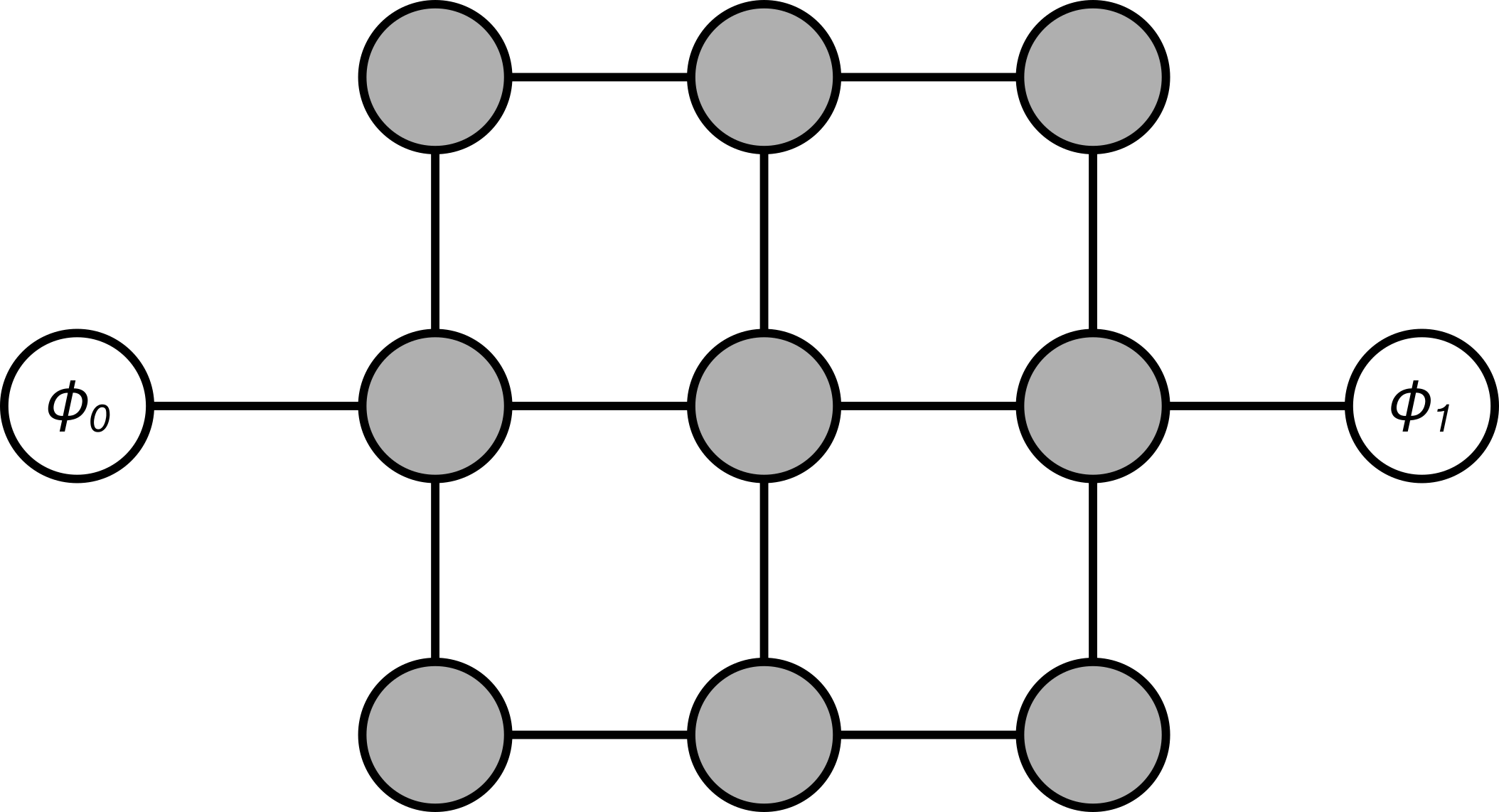}
    \caption{Lattice topology of the macrospin simulation. Each circle represents a macrospin. White circles ($\phi_0$ and $\phi_1$) are neuronal macrospins with easy-plane anisotropy and other relevant interactions; gray circles are synaptic macrospins. Connections between macrospins indicate the presence of an exchange interaction.}
    \label{fig:3x3-lattice}
\end{figure}

Note that the total angular momentum generated by a spike event is largely independent of the angle $\epsilon$. The notion that $\epsilon$ can be made arbitrarily small while the spiking energy remains roughly constant is a crucial feature of this system for neuromorphic application. Guarantees on stable spiking behavior near a small value of $\epsilon$ enable neural fan-out, as the large angular momentum burst of a single neural spike can be split up to trigger multiple $\epsilon$-thresholded neurons downstream, where the fan-out count depends on $\epsilon$ and the spike energy. To show that this is possible, we simulate a small $3\times 3$ square lattice of weakly easy-axis macrospins to which is attached two ``neuron'' macrospins that nominally obey Eq.~\eqref{eq:phi-ddot}. The setup is depicted in Fig.~\ref{fig:3x3-lattice}.

\begin{figure}
    \centering
    \includegraphics[width=\columnwidth]{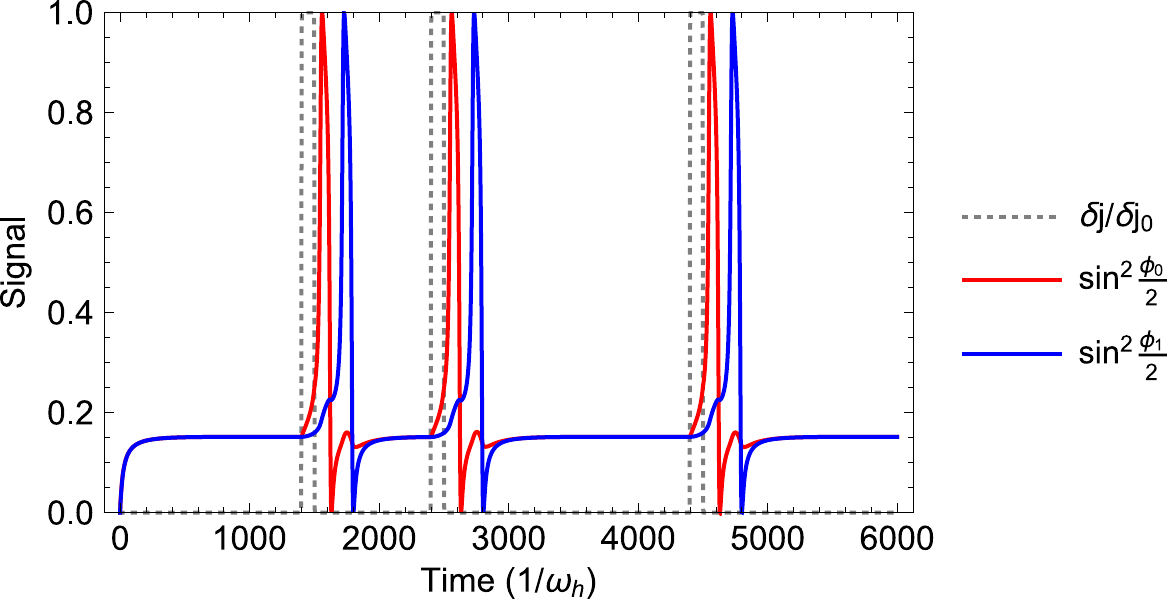}
    \caption{Spiking dynamics of the in-plane phase for the neuronal spins $\phi_0$ and $\phi_1$. The system is held close to threshold until a small current pulse is applied to $\phi_0$ during a time interval during indicated by the dotted lines. This causes a phase spike for $\phi_0$ (red curve). A short time later, $\phi_1$ spikes due to excitations from the $\phi_0$ spike pushing it over its threshold (blue curve). The process is then repeated multiple times with different intervals.}
    \label{fig:1to1-spikes}
\end{figure}

To ensure that the neuronal spikes undergo $2\pi$ phase slips and don't get stuck on the easy axis potential well at $\phi=\pi$, we apply a weak magnetic field $\bm B = B\bm e_x$ to the entire system. This modifies Eq.~\eqref{eq:phi-ddot} through the addition of a $B\sin\phi$ term on the left-hand side. We tune the current $j$ so that the system sits just below the autooscillation regime, and then apply a small perturbative current $\delta j$ (around 3.5~\% of $j$) over a small time window $\Delta t$ shown by dotted lines in Fig.~\ref{fig:1to1-spikes}. We find that this generates a spike (red curve in Fig.~\ref{fig:1to1-spikes}, and the angular mometum of that spike deposited into the lattice is strong enough to trigger spiking events in the other neuron as well (blue curve in Fig.~\ref{fig:1to1-spikes}. 
Note that $\delta j$ is turned off well before the spike peaks; it contributes just enough torque to push the neuron past its threshold. The rest of the energy for generating the spike is sourced from the global $j$ running through the entire system. In this way, $j$ acts as a local energy source for the neurons, similar to electrochemical energy stored in ATP in the brain or the voltage rails $V = 0$ and $V= V_\text{dd}$ in a CMOS circuit.

One also observes small bumps preceding the $\phi_1$ spikes in Fig.~\ref{fig:1to1-spikes}. Just as $\delta j$ triggers the spiking of $\phi_0$, these bumps trigger the spiking of $\phi_1$. A more detailed analysis reveals that these bumps correspond to a spin-$\hat y$ torque injected from the neighboring easy-axis macrospins in the lattice. Plots demonstrating this behavior, as well as details of the macrospin model, are presented in Appendix~\ref{app:macrospin-details}.

The macrospin model illustrates the principles of operation we envision for a neuromorphic system built on easy-plane spiking oscillators, but it is limited by the scope of its realism. In the remainder of the paper, we explore the neuromorphic system in micromagnetic simulations. Although the microscale physics becomes significantly more complex, we recover the same neural behavior, and investigate additional interesting phenomenon such as a multi-neuron chains, synaptic weights, and additive fan-in. 

\section{Micromagnetic simulations}

\begin{figure}
\includegraphics[width=\columnwidth]{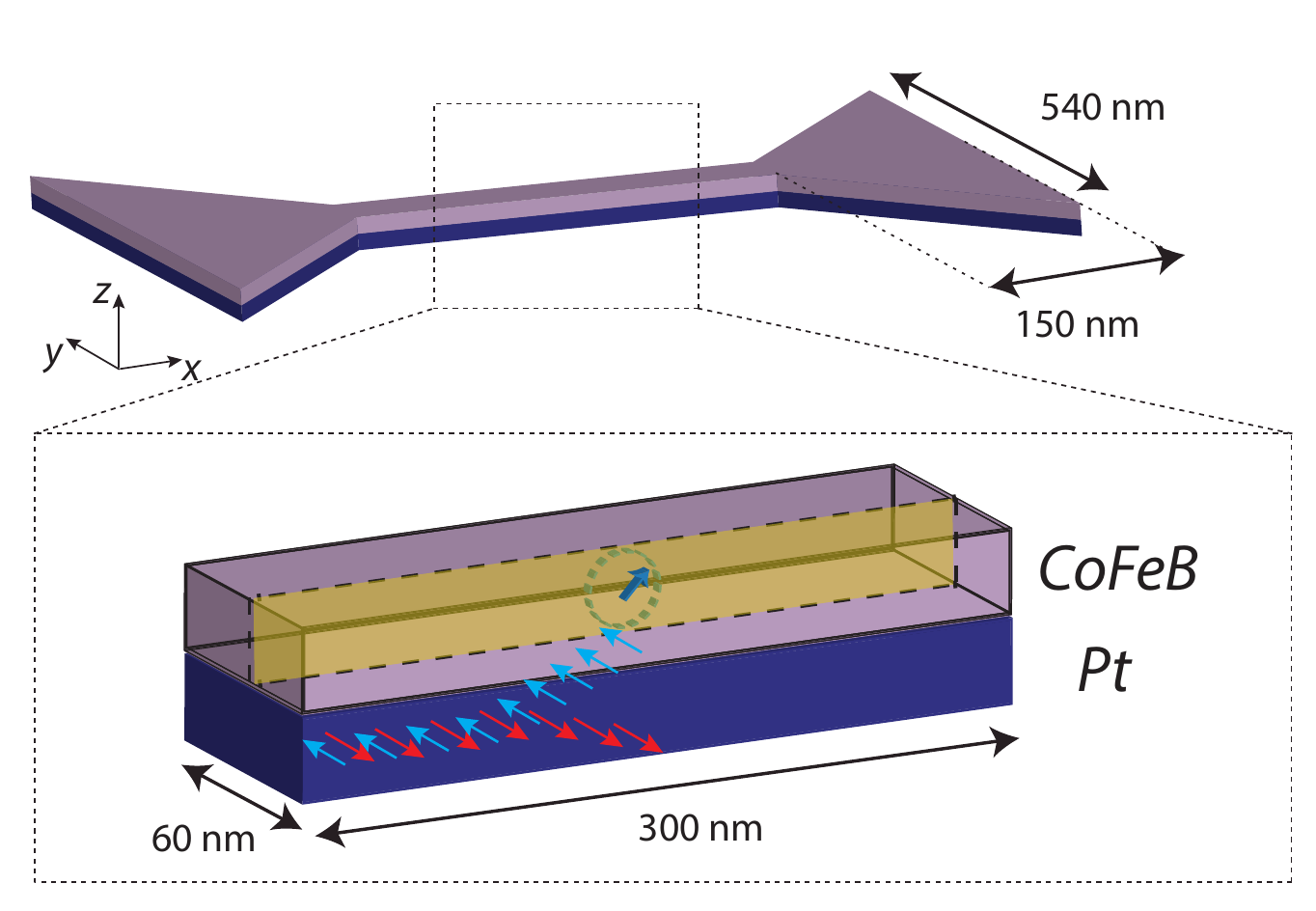}
\caption{Geometry for micromagnetic simulations. The bottom layer is normal heavy metal such as Pt and the top layer is a 2~nm thick CoFeB. The dark blue arrow shows large in plane (yellow) 2$\pi$ precessions of the top layer magnetization.}
\label{nanoconstriction}
\end{figure}
We perform micromagnetic simulations to show that easy-plane oscillations and voltage spikes can be obtained in a nano-constriction oscillator architecture that is convenient for coupling oscillators in chains~\cite{Zahedinejad2019}. We consider the geometry shown schematically in Figure~\ref{nanoconstriction}. The nano-constriction is 300~nm long and 60~nm wide with arms that are 150~nm long and 540~nm wide.

\begin{figure}
\includegraphics[width=\columnwidth]{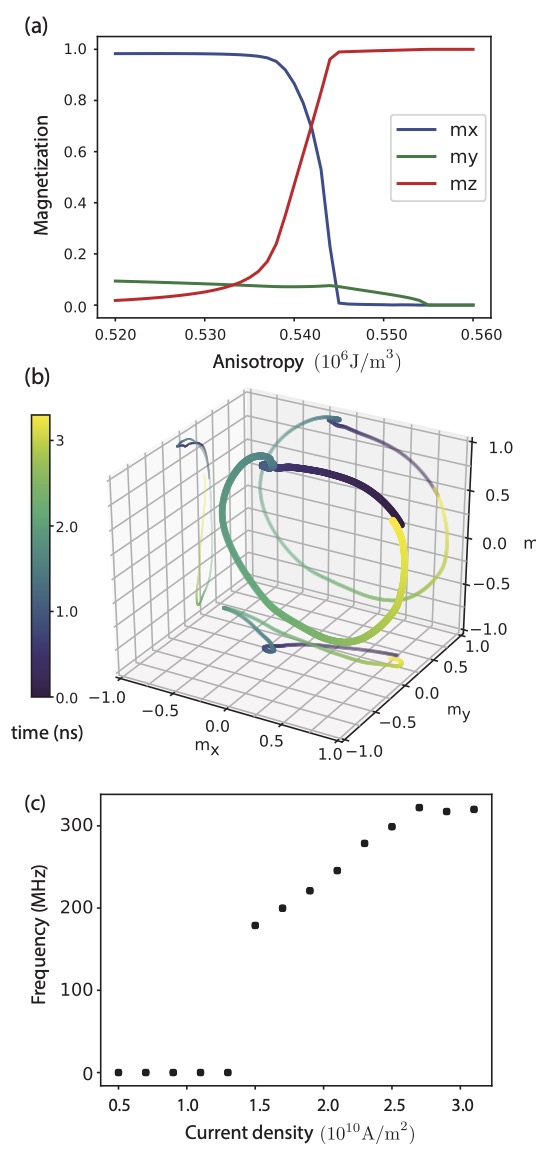}
\caption{(a) Scaled magnetization components as a function of perpendicular magnetic anisotropy $K_{\text{u}}$. Transition from in-plane ($m_x = 1$) to out-of-plane ($m_z = 1$) magnetization is observed for $K_{\text{u}} = 0.54 \times 10^6$ J/m$^3$. We use \textit{minimize} routine from MuMax3~\cite{Vansteenkiste2014} that finds the ground state using the conjugate gradient method. (b) Magnetization precession in the $x-z$ plane with color coded time for input current density j = 2.4 $\times 10^{10}$ A/m$^2$. Discrepancies from the circular trajectory due to the exact nano-constriction shape are visible in the projections on the three planes. (c) Oscillation frequency as a function of the current density. Critical current density is 1.5 $\times 10^{10}$ A/m$^2$.}
\label{easy-plane}
\end{figure}

Easy-plane dynamics requires a ferromagnetic material with an easy-plane perpendicular to the polarization of the spin current. Previous studies have used spin valves with perpendicular spin current polarizer, which has the benefit of easy-plane in the magnetic film plane that is easy to obtain, but at the price of a  complicated magnetic film stack fabrication~\cite{Houssameddine2007,Ebels2008,Rowlands2019}. Here we make the choice to use the spin Hall effect to inject a spin current. In this case, fabrication is simple, but the spin current polarization is in the film plane, such that the easy-plane needs to be perpendicular to the film plane. 
\begin{figure}
\includegraphics[width=\columnwidth]{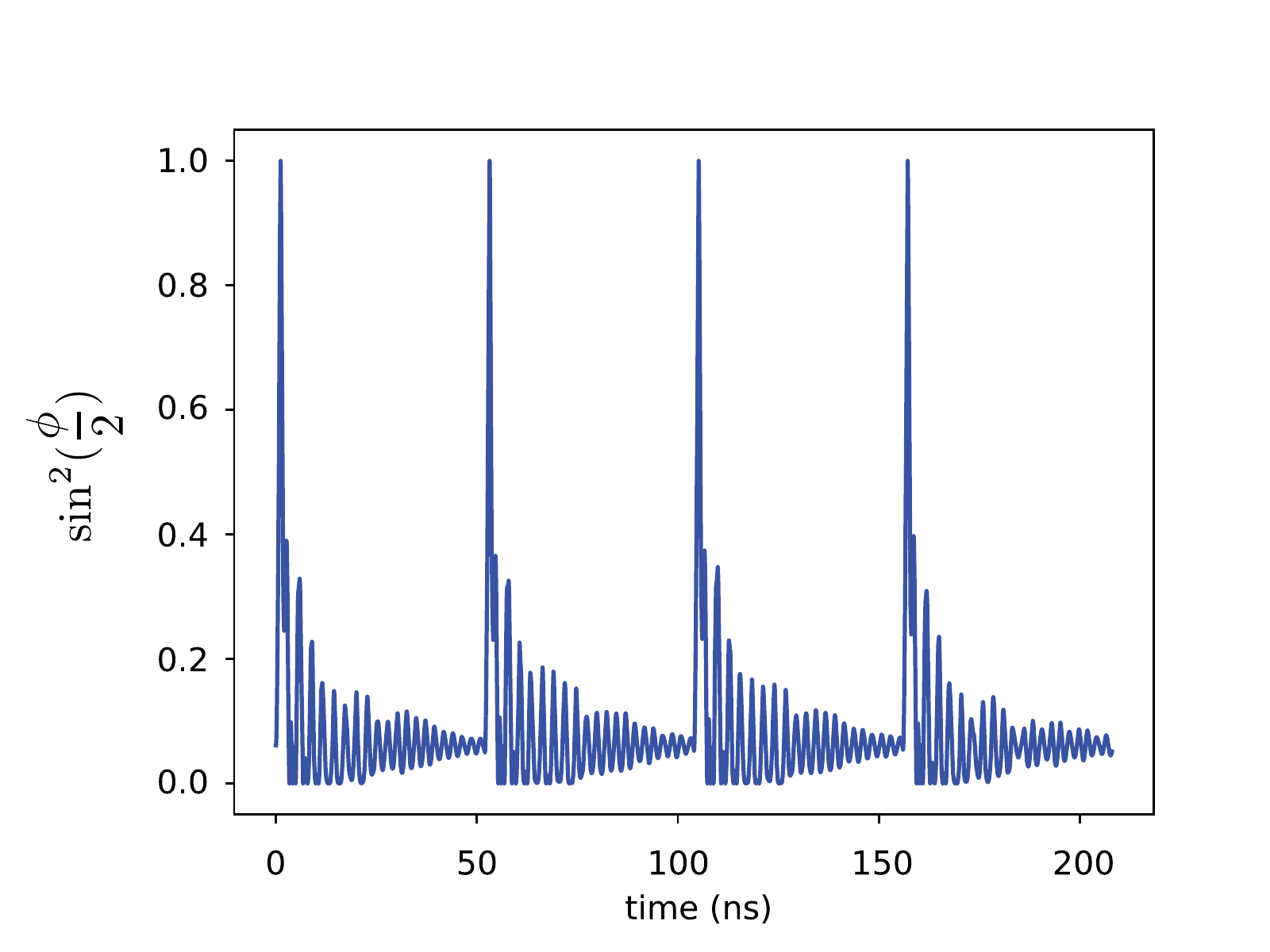}
\caption{Normalized tunneling magnetoresistance of the nano-constriction easy-plane SHNO. Current pulses with density $j = 2.5 \times 10^{10}$~A/m$^2$ are applied over 2~ns time intervals spaced by 50~ns. }
\label{spikes}
\end{figure}

To obtain the perpendicular easy-plane we choose the strategy that was already used to obtain magnetic skyrmions in Co, that relies on adjusting the ferromagnet thickness such that the demagnetizing field compensates the PMA~\cite{Legrand2019}. In the simulations we fix the thickness and determine the average equilibrium magnetization along the different axis for different values of PMA (Fig.~\ref{easy-plane} (a)). We consider 2~nm thick CoFeB with saturation magnetization $M_{\text{s}} = 9.55\times10^5$ A/m, exchange constant $A_{\text{ex}} = 2 \times 10^{11}$~J/m and magnetic damping $\alpha = 0.01$. We find the easy plane anisotropy at the transition from in plane (m$_x$=1) to out of plane (m$_z$=1) magnetization for the z-axis anisotropy of $K_{\text{u}} = 0.54 \times 10^6$ J/m$^3$. 

In the following, we set the anisotropy to the compensation value and we apply a spin current $\vec{j}_{\text{s}} = j_{\text{s}} \times \hat{e}_y $ with the spin polarization set to 1. For a current density of $1.5 \times 10^{10}$~A/m$^2$ we obtain large circular oscillations in the perpendicular easy plane that present distortions compared to the simplified macrospin model, as can be observed in Fig.~\ref{nanoconstriction}(b) ~\cite{Timopheev2016}. This spin current density is comparable to those typically found in spin-torque nano-oscillators~\cite{Dussaux2010a}, making it realistic for experimental realizations.

We find easy-plane dynamics for nano-constriction widths of 20~nm, 40~nm, and 60~nm, but not for larger widths. This is due to the fact that magnetization confinement is not strong enough for larger widths and the hard-axis anisotropy is not large enough. For the 60~nm width the precession orbit is less circular than for smaller widths; nevertheless, all the results we report in this paper use the 60~nm width because of the greater ease of device fabrication with e-beam lithography. Furthermore, we find that it is important that $y$ dimensions of the arms are larger than their $x$ dimensions, such that their magnetizations are aligned along the $y$ axis and thus do not interfere with the $xz$-plane oscillations in the nano-constriction. 

\begin{figure*}
\includegraphics[width=\textwidth]{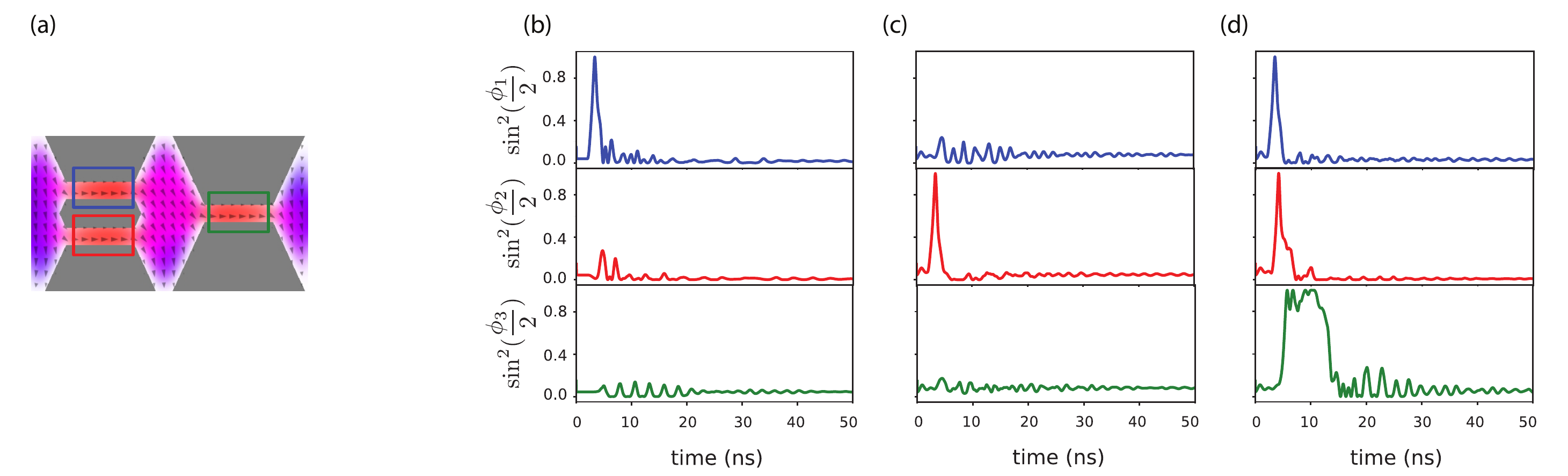}
\caption{(a) Top view of a ``two-to-one'' neural network building block. (b) Current input only in the first of the input layer neurons (blue), the output neuron (green) does not spike. (c) Current input only in the second of the input layer neurons (red), the output neuron does not spike. (d) When both input neurons spike, the energies of output spin waves sum up and make the output layer neuron spike.}
\label{two_inputs}
\end{figure*}

We use Hilbert transforms to extract the instantaneous oscillation frequency as a function of current density (Fig.~\ref{easy-plane}(c)). The frequency varies approximately linearly over a range of $1 \times 10^{10}$~A/m$^2$. For increasing current density the magnetization cants more and more out of the easy-plane. For $j = 2.7 \times 10^{10}$~A/m$^2$ the trajectory starts to deviate considerably from a circle in the easy-plane, which makes the frequency saturate. At $j = 3.3 \times 10^{10}$~A/m$^2$ the out-of-plane angle becomes too large and there are no more coherent oscillations. 

In simple bilayer spin Hall structures, magnetization dynamics can be detected through the anisotropic magnetoresistance effect~\cite{Wang2002}, arising from the fact that the resistance of the device is dependent on the mutual orientation between the electric current and the magnetization. Alternatively, a tunnel junction could be added on the top such that the magnetization dynamics could be detected through the tunneling magnetoresistance effect which has a larger amplitude. For magnetization states confined to the easy plane 
\begin{equation}
    R(\phi)-R(0) \sim \sin^2\left(\frac{\phi}{2}\right) .
\end{equation}
In the figures below, we plot the angular dependence (rather than the resistance directly) to maintain a close analogy with the physics of the Josephson junction phase.
To obtain the voltage spikes, we apply a current pulse for a duration that is equal to a single oscillation period, on the order of 2~ns for the geometrical parameters we chose. The magnetization makes a single turn corresponding to a single spike in magnetoresistance and then slowly relaxes towards the easy axis within the easy plane. It takes about 50~ns to completely relax. After this relaxation time, the nano-constriction oscillator neuron can be excited and spike again, as shown in Fig.~\ref{spikes}.

Easy-plane nano-constriction spin Hall oscillators can thus emulate the spiking behavior of biological neurons. In the following section we begin the exploration of how they can be assembled in physical neural networks and used to encode and process information through dynamics.

\section{Proposal for neuromorphic computing - Spike propagation in chains of nano-constrictions}

Easy-plane spin Hall neurons can be assembled in a neural network as a two-dimensional array of nano-constrictions. In this section we simulate two building blocks of such a neural network that implement operations essential for neuromorphic computing.

First we show that outputs from two neurons in the same neural network layer can be summed and injected into a neuron in the following layer. For this we simulate the architecture shown in Fig.~\ref{two_inputs}(a). The two nano-constrictions on the left correspond to the two input layer neurons, and the nano-constriction on the right to a single output layer neuron. Input neurons can receive an input in the form of a current pulse. In order to apply current pulses, we envision placing electrodes on each island and using additive and subtractive currents to achieve placing the desired current through a single nano-constriction, as in Ref.~\cite{Romera2017a}. We show that if only one of the input neurons spikes (Fig.~\ref{two_inputs}(b--c)), the energy transmitted to the output neuron by the spin waves propagating in the island between the nano-constrictions is not sufficient to make it spike. However if both input neurons spike at the same time, the output neuron spikes as well (Fig.~\ref{two_inputs}(d)). In the simulations, a constant bias is applied in all the neurons at current density $j_{\text{bias}}=0.6 \times 10^{10}$~A/m$^2$, that is below the critical current density of $1.9 \times 10^{10}$~A/m$^2$ of the three nano-constrictions structure. The spike in the first (resp. second) neuron is induced by a 2.7 ns long input current pulse applied 2 ns (resp. 2.8 ns) after the beginning of the simulation.

\begin{figure*}
\includegraphics[width=\textwidth]{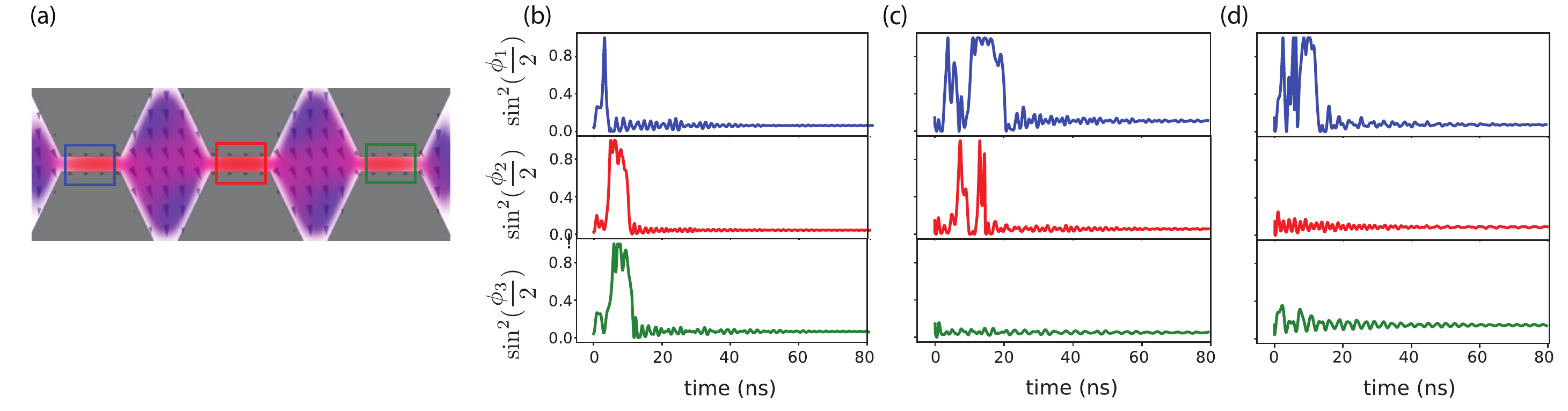}
\caption{(a) Top view of a chain of three nano-constriction neurons. (b) Responses of the three nano-constrictions show spike propagation. (c) When the anisotropy is modified in the second synapse, the first and second neurons spike, but the third one does not. (d) When the anisotropy is modified in the first synapse, the spikes do not propagate further. The first neuron spikes, but the two following ones do not.}
\label{spike_prop}
\end{figure*}

Second, we show that neural outputs can be multiplied by synaptic weights. To do this, we simulate a chain of three nano-constriction neurons, whose top view is shown in Figure \ref{spike_prop} (a). Only the first neuron in the chain receives the input current pulse. The whole chain is biased with a constant current $j_{\text{bias}}=1.6 \times 10^{10}$~A/m$^2$. An input spin current pulse with density $j_{\text{pulse}}=4 \times 10^{10}$~A/m$^2$ is applied to the first nano-constriction for a duration of 2.7~ns. The excitation from the first neuron propagates down the chain and with approximately 2~ns delays we observe spikes in the second and third junctions, see Figure \ref{spike_prop} (b). We demonstrate synaptic functionality by modifying the coupling between the nano-oscillators and thus controllably altering the spike propagation in the chain. This can be done by locally modifying magnetic properties such as anisotropy or damping in the synaptic islands between each pair of nano-constriction neurons. Indeed, neurons are magnetically coupled by spin waves that propagate in these islands. A discontinuity in magnetic properties induces spin wave reflections and lowers the coupling between nano-constriction oscillators. Here we choose to modify anisotropy rather than damping because experimentally it can be done \textit{in situ} by the application of a dc voltage~\cite{Zahedinejad2020}. Alternatively, damping could be modified by ion irradiation~\cite{Jiang2020} once the neural network has been trained offline and the physical neural network is prepared for inference.

The anisotropy in the whole magnetic structure is $K_{\text{u}} = 0.54 \times 10^6$ J/m$^3$. We locally modify the anisotropy by 10~\% in a 150 nm wide area in the center of the island between the second and third neuron, such that synaptic anisotropy becomes $K_{\text{u}}^{\text{syn}} = 0.6 \times 10^6$ J/m$^3$, and we observe that the first two neurons undergo spike bursts containing two spikes each, while the third neuron does not spike (Figure \ref{spike_prop} (c)). As the input current pulse is  of the same amplitude and duration, we interpret this emergence of spike bursts as due to the conservation of the total energy in the system. For the first neuron, the second spike lasts longer because the magnetization decelerates when aligned with the easy-axis as it is not receiving any drive current any more.

Similarly, when we increase the anisotropy of the first island, only the first neuron undergoes spike burst where its magnetization makes three whole turns in the easy plane, while the two following neurons stay still (Figure \ref{spike_prop} (d)).

We can thus decrease the synaptic weights by locally increasing the anisotropy in the islands by for example applying a dc voltage on this area. This breaks the magnetization dynamics in the island and thus impacts the mutual oscillator coupling.

\section{Conclusion}

\begin{figure}
    \centering
    \includegraphics[width=0.8\columnwidth]{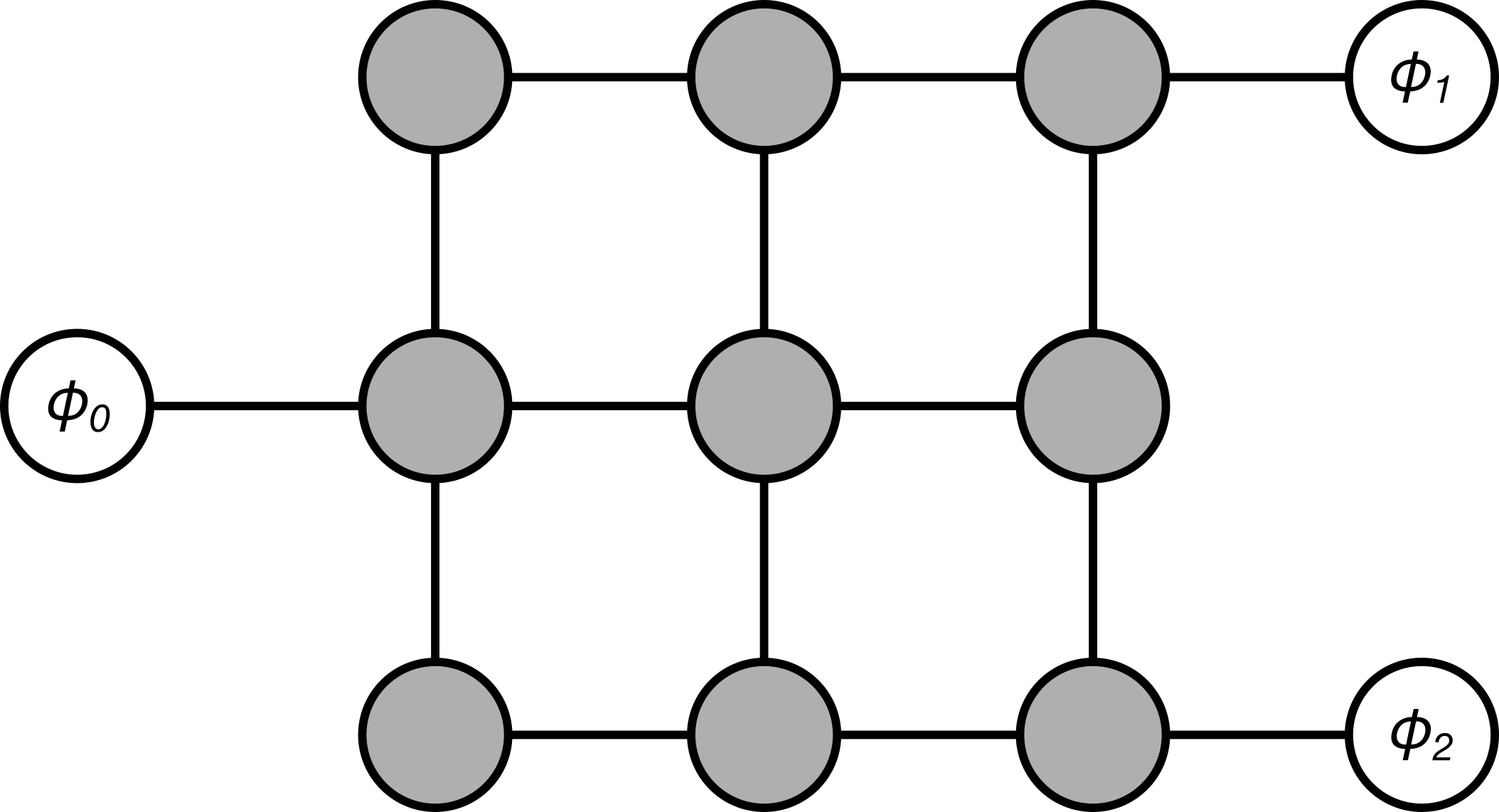}
    \caption{Lattice topology of the macrospin fan-out simulation. The ``upstream'' neuron $\phi_0$ generates a spike after being given a perturbative current $j$ on top of the static current $j$. The ``downstream'' neurons $\phi_1$ and $\phi_2$ both spike in response, as a consequence of spin excitations mediated by the $3\times 3$ lattice.}
    \label{fig:3x3-lattice-1to2}
\end{figure}

There are some attractive features of this approach to neurons and synapses. One is that the close-to-threshold current going through all of the neurons allows for gain, serving as a local energy source for generated spikes. That the spike energy is independent of how close to threshold the system sits could enable fan out.  We demonstrate the potential for fan-out behavior in a macrospin simulation. We extend the macrospin model of Sec.~\ref{sec:macrospin} (also detailed in Appendix~\ref{app:macrospin-details}) by including two downstream neurons rather than one. The lattice topology of the coupled macrospins is indicated in Fig.~\ref{fig:3x3-lattice-1to2}.

\begin{figure}
    \centering
    \includegraphics[width=\columnwidth]{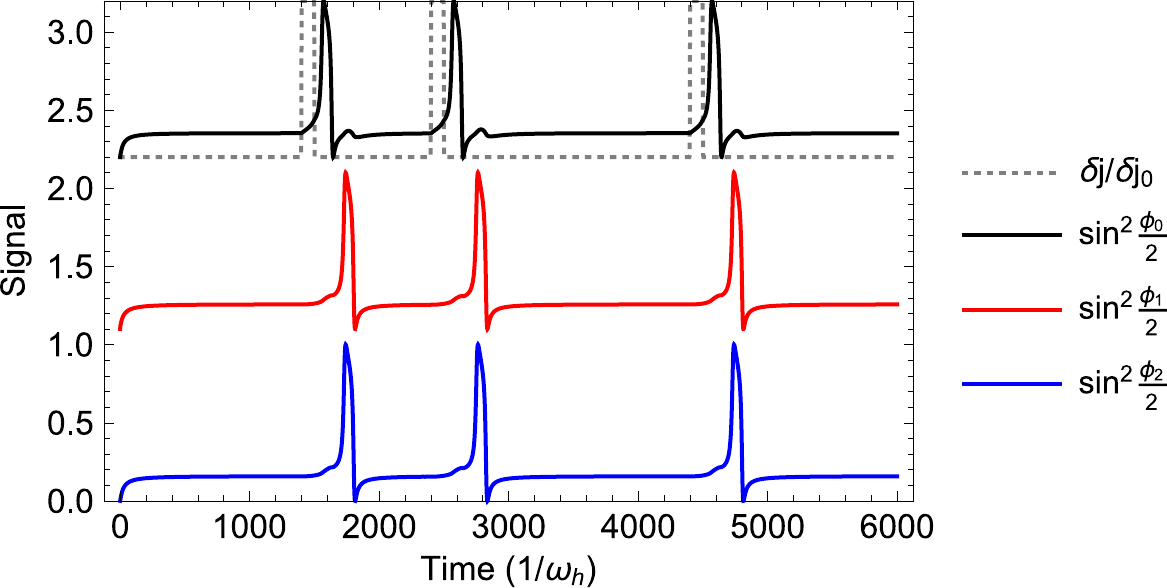}
    \caption{Demonstration of spiking fan-out in a simulation of coupled macrospins. A current $j$ running through the entire system keeps ``neuron'' macrospins close to their spiking threshold. Then a small perturbative current (dashed gray line) is applied \emph{only} to one macrospin (corresponding to $\phi_0$, the black line). This extra torque pushes the spin over its threshold, and it undergoes a single spiking event. Angular momentum propagates through the lattice (Fig.~\ref{fig:3x3-lattice}) and triggers spikes in both $\phi_1$ (red) and $\phi_2$ (blue). The plots are shifted arbitrarily in the vertical direction for clarity.}
    \label{fig:fan-out-spikes}
\end{figure}

Again we hold the static torque as high as possible, in this case $\tau = 0.288$, and again weakly coupled them $\omega_J = 0.3$ compared to the hard axis energy $\omega_h = 1$. All other parameters are the same as in Appendix~\ref{app:macrospin-details}. Just as before, we apply a small perturbative torque $\delta\tau = 0.001$ \emph{only} to the upstream neuron $\phi_0$. The excitation in the synaptic lattice generated by the spike on $\phi_0$ causes spiking activity in $\phi_1$ and $\phi_2$, shown in Fig.~\ref{fig:fan-out-spikes}.

In such a small macrospin system, the simplified dynamical mechanisms can sometimes lead to different essential physics than what one finds in micromagnetic simulations. We can verify by eye in the micromagnetic simulations that it is a pulse of spin wave energy that travels from one neuron to another and causes the target neuron to spike. But in the macrospin model, there exist regimes that transmit spikes not by spin wave but by effective single-domain switching driven by $\phi_0$.

\begin{figure}
    \centering
    \includegraphics[width=\columnwidth]{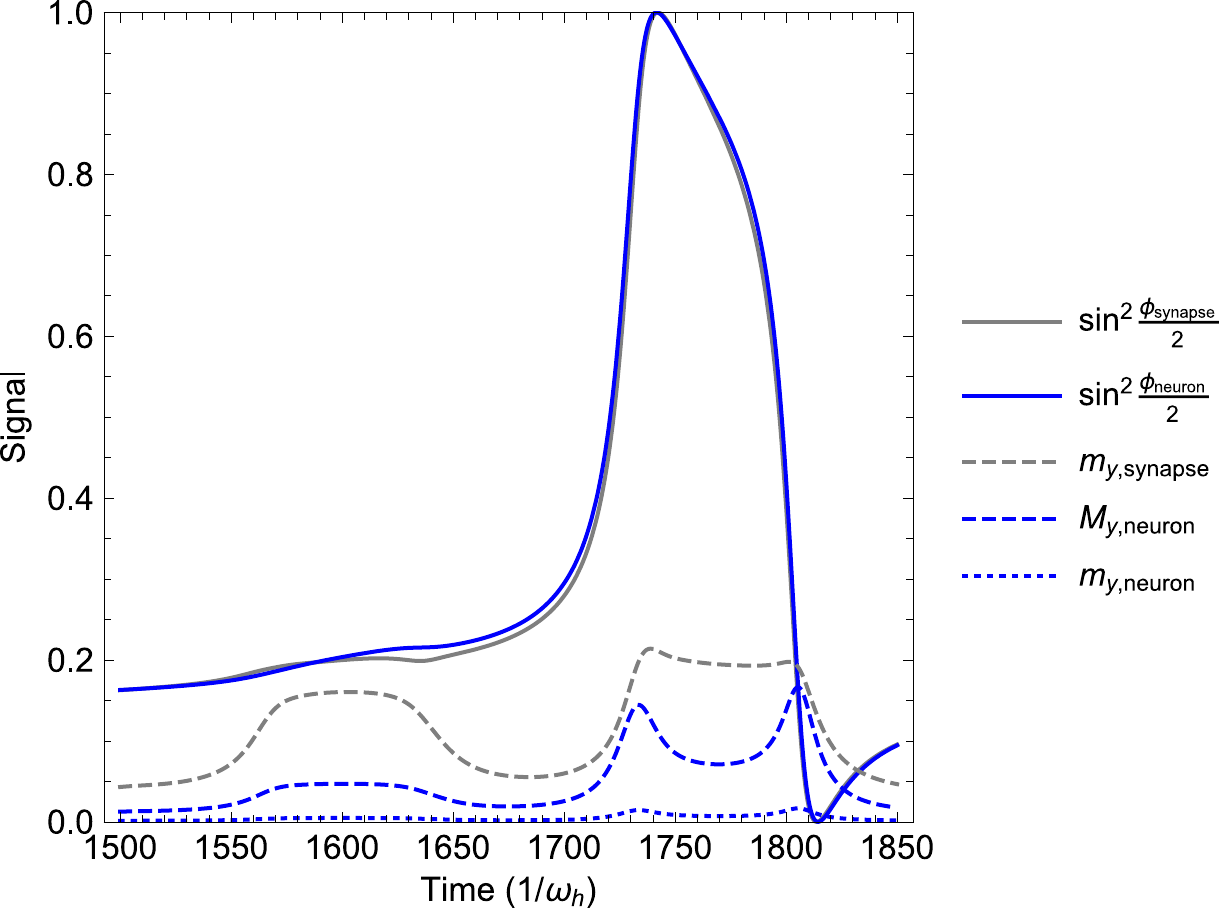}
    \caption{Spiking event interaction between a targeted neural spin ($\phi_\text{target}$) such as $\phi_2$ in Fig.~\ref{fig:3x3-lattice-1to2} and its neighboring spin in the lattice ($\phi_\text{synapse})$. By around $\omega_h t = 1550$, the influence of an upstream spiking event (from $\phi_1$) causes $\phi_\text{synapse}$ to cant in the $y$-direction. Being exchange coupled to $\phi_\text{neuron}$, the neural spin picks up a weak $y$-component and begins to precess, driven by the hard-axis field. Even once the $y$-torque from the synapse dies down (dashed gray line returning to equilibrium at $\omega_ht\approx 1675$, the neuron has already passed its threshold, and starts to spike. Note that $m_{y,\text{neuron}}$ is the normalized magnetization of the neuron, but in the simulation its non-normalized magnetization $M_{y,\text{neuron}}$ is ten times that of the synaptic spins.}
    \label{fig:causation}
\end{figure}

To verify that we are in a spin-wave driven regime, rather than a single-domain-driven one, we zoom in on one of the induced spiking events. Fig.~\ref{fig:causation} shows the angular behavior of an induced downstream neural macrospin in blue, and its neighboring lattice macrospin in gray. A wide packet of $y$-angular momentum (gray dashed line) arrives from the lattice centered around time $\omega_ht = 1600$. The injection of this angular momentum from synapse to neuron causes the neuron moment to tilt slightly out of plane, which in turns causes $\phi_n$ to precess slightly in the easy-plane anisotropy field. By the time the packet of angular momentum is used up (around $\omega_ht = 1600$), the neuron has already passed its spiking threshold.

Notice that as the spiking grows, the neural phase is advanced relative to the synaptic phase. This indicates that the spiking mechanism does \emph{not} arise from the lattice mimicking the source neuron in a single domain fashion and simply dragging the downstream neuron along with it. Rather, a small packet of momentum triggers a downstream spike event, and \emph{then} the nearby lattice spins are dragged through a spiking motion by this newly-spiking downstream neuron.
  
In conclusion, we have proposed a new geometry for an easy-plane ferromagnetic spin Hall oscillator. We have shown using a macrospin model that such oscillator can produce voltage spikes and thus emulate a biological neuron. We have then shown in micromagnetic simulations that such easy-plane geometry can be obtained in compensated PMA ferromagnets and that it can be conserved by fabricating oscillators in the nano-constriction geometry which is very convenient for coupling the oscillators in chains. Finally, we show that spike propagation between the nano-constrictions can be controlled, thus giving a  proof-of-principle demonstration of synaptic functionalities.

This research was supported by the Quantum Materials for Energy Efficient Neuromorphic
Computing (Q-MEEN-C), an Energy Frontier Research Center funded by the U.S. Department of Energy (DOE), Office of Science, Basic Energy Sciences (BES), under Award
DE-SC0019273.

\appendix 

\section{Macrospin model details}
\label{app:macrospin-details}
In Sec.~\ref{sec:macrospin}, we used a simple macrospin model to demonstrate spiking and fan-out behavior. In this Appendix, we clarify details of that model, and show that the downstream spiking is in fact driven by a small torque that surpasses a neural threshold rather than any sort of  single-domain spiking behavior that could plausibly occur in such a small, simple model.

We simulate the model using the eleven spins connected to their nearest neighbors by a simple exchange interaction $J$ in the topology of Fig.~\ref{fig:3x3-lattice}. It is convenient to work in dimensionless units in which all frequencies are scaled by $\omega_h$ (time scaled by $1/\omega_h$) to set the timescale similar to that of the spiking behavior. All quantities reported below are dimensionless. We weakly couple the spins by setting the characteristic exchange frequency $\omega_J = 0.31$, which is necessary to avoid a spiking event simply driving single domain switching via strong exchange coupling. We use a large value $\alpha = 0.5$ of the Gilbert damping to represent dissipation modes that cannot be modeled in the macrospin system, which are necessary to ensure an overdamped return to the balance point $\phi^*$ as outlined in Sec.~\ref{sec:macrospin}. The neuron macrospins are subject to a weak easy axis anisotropy of characteristic frequency $\omega_e = 0.05$ along $\hat x$, while the synaptic lattice macrospins have an out of plane easy axis $\omega_z = 0.2$ and a weak easy axis $\omega_s = 0.1$ along $\hat y$ that models the shape anisotropy of the micromagnetic system. A weak magnetic field with characteristic frequency $\omega_B = 0.05$ is applied to all spins to destabilize the $\phi = \pi$ potential well. Finally, we artificially inflate the saturation magnetization of the neuronal spins by a factor of ten compared to the lattice spins. This models the comparative softness of the synaptic modes compared to the hard neuronal modes observed in micromagnetics.

The analysis of Sec.~\ref{sec:macrospin} suggests that the $\pi/4$ threshold sits at $\tau = \omega_e/2 = 0.025$. We find that we can push $\tau$ slightly higher than this, to $\tau = 0.02895$, due to applied $B$ field. At selected times, we add a spike of height $\delta\tau = 0.001$. This generates the neuronal spikes observed in Fig.~\ref{fig:1to1-spikes}.

\end{document}